\documentclass[twocolumn,aps,pra,amssymb,amsmath,%
    superscriptaddress]{revtex4}

\newcommand*{\id}{\openone}
\newcommand*{\ket}[1]{|{#1}\rangle}
\newcommand*{\bra}[1]{\langle{#1}|}
\newcommand*{\braket}[2]{\langle{#1}|{#2}\rangle}
\newcommand*{\HH}{{\mathcal H}}  
\newcommand*{\EE}{{\mathcal E}}  
\newcommand*{\eref}[1]{Eq.~(\ref{#1})}
\newcommand*{\Eref}[1]{Eq.~(\ref{#1})}
\DeclareMathOperator{\diag}{diag}

\begin{document}
\title{General criterion for oblivious remote state preparation}
\author{Z. Kurucz}
\affiliation{Department of Nonlinear and Quantum Optics, Research
  Institute for Solid State Physics and Optics, Hungarian Academy of
  Sciences, P.O. Box 49 H-1525 Budapest, Hungary}
\author{P. Adam}
\author{J. Janszky}
\affiliation{Department of Nonlinear and Quantum Optics, Research
  Institute for Solid State Physics and Optics, Hungarian Academy of
  Sciences, P.O. Box 49 H-1525 Budapest, Hungary}
\affiliation{Research Group for Nonlinear and Quantum Optics, Hungarian Academy of
  Sciences and Institute of Physics, University of P\'ecs,
  Ifj\'us\'ag \'ut 6. H-7624 P\'ecs, Hungary}

\date{8 June, 2005}

\begin{abstract}
  A necessary and sufficient condition is given for general exact
  remote state preparation (RSP) protocols to be oblivious, that is,
  no information about the target state can be retrieved from the
  classical message.  A novel criterion in terms of commutation
  relations is also derived for the existence of deterministic exact
  protocols in which Alice's measurement eigenstates are related to
  each other by fixed linear operators similar to Bob's unitaries.  For
  non-maximally entangled resources, it provides an easy way to search
  for RSP protocols.  As an example, we show how to reduce the case of
  partially entangled resources to that of maximally entangled ones,
  and we construct RSP protocols exploiting the structure of the
  irreducible representations of Abelian groups.
\end{abstract}
\pacs{03.67.-a}%
\keywords{quantum communication,
  exact oblivious remote state preparation,
  entanglement, teleportation, antilinear operators}
\maketitle

\section{Introduction}
\label{sec:intro}

Quantum communication, i.e., the transfer of quantum information can
involve three kinds of resources: entanglement, classical
communication, and direct transfer of quantum information.  Quantum
communication protocols that make use of the entangled resource
consist of a suitable measurement performed on the sender's half of
the entangled pair, followed by classical communication.  The
measurement depends on the target state that is to be transmitted, and
the classical message respects the measurement outcome.  Essentially,
this action can always be viewed as a remote state preparation (RSP)
scheme \cite{pra63e014302, pra69e022310, pra65e022316, prl90e127905,
  pra67e052302, prl87e077902, tit51p56, pra72e052315, pra62e013802,
  ijqi1p301, prl87e197901, pra68e062319, pra70e062306}.  The goal of
RSP is to prepare a given quantum state at a distant location using
only shared entangled pairs and classical communication as resources.

An intriguing question about RSP is whether the relevant information
describing the target state is transmitted over the classical
communication channel \cite{prl87e197901, pra68e062319, pra70e062306}
or the entangled resource \cite{pra63e014302, prl90e127905,
  pra65e022316, pra67e052302, pra69e022310, prl87e077902, tit51p56,
  pra72e052315, pra62e013802}.  In the latter case, no access to it
can be gained from the classical message.  Such protocols are called
\emph{oblivious}.  This aspect is of crucial importance if, e.g.,
quantum communication is introduced for securing a private
communication channel.  Therefore, it is worth to investigate in what
conditions an RSP protocol is oblivious.  In this Paper, we give a
necessary and sufficient condition for a generic exact RSP protocol to
be oblivious.  A protocol is called exact (or faithful) if it succeeds
with unit fidelity.  For the existence of those deterministic exact
RSP protocols in which the measurement eigenstates are related to each
other by linear operators similar to Bob's unitaries, we provide a
simple condition in terms of commutation relations.  We give a method
for constructing such protocols and trace them back to the case of
maximally entangled resources.

The paper is organized as follows.  In Sec.~\ref{sec:antilinRSP}, we
introduce a mathematical model of RSP and argue for the importance of
antilinear operators in RSP.  In Sec.~\ref{sec:oblivRSP}, we prove the
necessary and sufficient condition of obliviousness.  Then we
investigate exact deterministic protocols using partially entangled
states and give an example how to reduce them to simpler ones in
Sec.~\ref{sec:exact}.  Finally, Sec.~\ref{sec:summary} summaries our
results.

\section{Antilinearity involved in RSP}
\label{sec:antilinRSP}

We start from an RSP protocol in which the sender (Alice) and the
receiver (Bob) share a pure but not necessarily maximally entangled
state $\ket\Psi_{AB}$.  The use of mixed resources would prevent the
scheme from being able to exactly prepare a pure state.  Suppose that
Alice performs a von Neumann measurement whose result is given by
the rank one projection $\ket\phi_{AA}\bra\phi$.  Alternatively, it
may be a rank one positive operator with $\|\phi\|^2<1$ for a POVM
measurement.  Due to the very nature of the von Neumann projection
principle, Bob's unnormalized conditional state is an antilinear
(conjugate linear) function $R \colon \HH_A \to \HH_B$ of Alice's
measurement eigenstate
\begin{equation}
  \label{eq:defRSO}
  \ket\psi_B \propto {}_A \braket\phi\Psi _{AB} 
  \equiv R\ket\phi_A ,
\end{equation}
where the partial scalar product is antilinear in its first argument,
and the norm square of the right hand side gives the probability of
this measurement event,
\begin{equation}
  \label{eq:p}
  p = \|{}_A\braket\phi\Psi_{AB}\|^2 = 
  \|R\ket\phi_A\|^2.
\end{equation}

The entangled state $\ket\Psi_{AB}$ is completely given by $R$ that is
mapping a possible measurement eigenstate in system $A$ to the
corresponding state of system $B$ after the measurement.  Formally, $R$
is nothing more than the composition of a complex conjugation (like time
inversion) with a linear operator, and it is diagonal in the Schmidt
basis.  The isomorphism between pure entangled states and antilinear
operators, and its application to quantum teleportation has been
thoroughly investigated in Refs.\ \cite{lnp539p93, fdp49p1019,
  job5pS627}.  Without loss of generality, we can assume that the
Hilbert spaces of Alice and Bob are both $D$ dimensional, and that the
shared entangled state has nonzero Schmidt coefficients only, that is,
the antilinear operator defined in (\ref{eq:defRSO}) is invertible.

A sufficient and necessary condition for the existence of exact
deterministic RSP protocols was given in Ref.~\cite{pra67e052302}.  It
was shown that a target state $\ket{\psi_0}_B$ can be prepared
unconditionally if and only if there is such a set of unitary
transformations $U_m$ under Bob's control and the probability
distribution $p_m(\psi_0)$ for Alice's measurement is such that
\begin{equation}
  \label{eq:usualRSPcond}
  \sum_{m=0}^{N-1} p_m(\psi_0)
  U_m^\dag \ket{\psi_0}_{BB} \bra{\psi_0}
  U_m = \rho_B,
\end{equation}
where $\rho_B$ is the partial density operator of the entangled state
$\ket\Psi_{AB}$ and $N$ is the number of possible outcomes. (For a
projective measurement $N=D$, while $N>D$ for a POVM.)
\Eref{eq:usualRSPcond} expresses the fact that the density operator of
system $B$ is not changed by Alice's action as long as the outcome of
the measurement is unspecified.  Then the eigenstates of Alice's
measurement can be constructed as
\begin{equation}
  \label{eq:usualPOVM}
  \ket{\phi_m}_A = \sqrt{p_m(\psi_0)} R^{-1} U_m^\dag \ket{\psi_0}_B,
\end{equation}
where $R$ is the antilinear operator defined in (\ref{eq:defRSO}) that
describes the entangled resources.  The set $\EE_B \subset \HH_B$ of
states is called a \emph{preparable ensemble} if it consists of states
that can be prepared using the same fixed set of unitaries $U_m$.

If the exact deterministic RSP protocol is oblivious---that is, the
probability distribution $p_m(\psi_0)$ is independent of the target
states $\ket{\psi_0}_B \in \EE_B$---then \eref{eq:usualPOVM} implies
that the eigenstates of Alice's measurement are antilinear functions
of the target state: for each successful outcome $m$, there exists an
operator $A_m \colon \HH_B\to\HH_A$ such that
\begin{equation}
  \label{eq:geneigenstates}
  \ket{\phi_m}_A = A_m \ket{\psi_0}_B,
  \quad
  A_m = \sqrt{p_m} R^{-1} U_m^\dag,
\end{equation}
and $A_m$ is antilinear and is the same for all $\ket{\psi_0}_B \in
\EE_B$.

Let us verify the existence of such antilinear operators in two
well-known examples.  Quantum teleportation of a qubit state can be
regarded as an oblivious RSP protocol.  Alice performs a POVM
measurement consisting of four one-rank positive operators.  One of
them is
\begin{equation}
  \label{eq:telepovm}
  \ket{\phi_0}_{AA} \bra{\phi_0} = \frac12 R^{-1} 
  \ket{\psi_0}_{BB} \bra{\psi_0} R^{-1\dag},
\end{equation}
with $\|\phi_0\|^2=1/2$ and $U_0 = \id_B$ being the identity operator.
The other three positive operators are $\ket{\phi_i}_{AA}
\bra{\phi_i}$ where the (unnormed) vectors $\ket{\phi_i}_A$ are
obtained from $\ket{\phi_0}_A$ by the Pauli matrices.  All the four
eigenstates are related to the target state in an antilinear way,
because $A_0 = \frac12 R^{-1}$ and $A_i = \frac12 \sigma_i R^{-1}$.

Another kind of oblivious RSP protocol was introduced in Ref.~\cite
{pra63e014302} that needs half the amount of classical communication
than teleportation.  In this scheme, Alice performs a von Neumann
projective measurement, one of its eigenstates being $\ket{\phi_0}_A =
\frac1{\sqrt2} R^{-1} \ket{\psi_0}_B$ with $\|\phi_0\|^2=1$.  The
other eigenstate $\ket{\phi_1}_A$ is orthogonal (antipodal) to
$\ket{\phi_0}_A$.  For states $\ket{\phi_0}_A$ chosen from a special
ensemble, namely, for states on a great circle of the Bloch sphere,
the $\pi$-rotation with axis perpendicular to the circle is the
unitary transformation $U$ that maps $\ket{\phi_0}_A$ to
$\ket{\phi_1}_A$.  Therefore, we have two antilinear operators $A_0 =
\frac1{\sqrt2} R^{-1}$ and $A_1 = \frac1{\sqrt2} U R^{-1}$, that are
antiunitary as well in this case.  Unlike in quantum teleportation,
the ensemble of preparable states $\EE_B$ is restricted to a great (or
small~\cite{pra69e022310}) circle in this protocol.

We note that the POVM operators, like in \eref{eq:telepovm}, are
linear functions of the matrix elements of the target density operator
$\ket{\psi_0}_{BB} \bra{\psi_0}$, since complex conjugation is
equivalent to transposition for Hermitian matrices.  The corresponding
linear superoperators are, however, not completely positive.

Though the eigenstates of the POVM operators are necessarily
antilinear functions of the target state, this fact does not exclude
the possibility that, in some special cases, we can find linear
operators $L_m \colon \HH_B\to\HH_A$ as well that also generate the
eigenstates from the target state, $\ket{\phi_m}_A = L_m
\ket{\psi_0}_B$.  This can happen, for example, in the trivial case
when $\EE_B$ consists of a basis $\ket i_B$ and nothing more.  Then we
can define the linear operators such that $L_m \ket i_B \equiv A_m
\ket i_B$.  However, this kind of ``classical RSP'' is nothing more
than a classical hashing protocol with entanglement providing the
shared secret key (c.f.\ one-time pad), and it is out of the scope of the
present Paper.

A similar construction of linear operators is also possible when
$\EE_B$ consists of real superpositions of the basis states.  On this
real subspace, complex conjugation does not play role, the antilinear
$A_m$ equals to the linear $L_m$, so antilinearity is not an issue.
However, as it is shown in Ref.~\cite{pra65e022316}, RSP is realizable
in real Hilbert space only if the dimension of the space is 2,~4,
or~8.  This means that $\EE_B$ is usually not a real subspace, so
complex conjugation does matter and antilinear functions are the
simplest choice to relate measurement eigenstates to the target state.

\section{Condition of obliviousness}
\label{sec:oblivRSP}

We have seen in Sec.~\ref{sec:antilinRSP} that a \emph {necessary
  condition} for an exact RSP scheme to be oblivious is that \emph
{for all the outcomes $m$, either there exists a fixed antilinear
  operator $A_m$ mapping the target state $\ket{\psi_0}_B$ to the
  measurement eigenstate $\ket{\phi_m}_A$, or Bob registers the
  outcome $m$ as a failure.}  The question naturally arises whether
the existence of such antilinear operators is a sufficient condition
of obliviousness.  In this section, we show that the answer is
positive.

Suppose that we are given an exact RSP protocol in which there exist
such antilinear operators $A_m$, but we do not require that the
protocol is oblivious.  We utilize the result of~\cite{pra67e052302}
that it suffices for Bob to perform a unitary operation in order to
restore the target state.  (Actually, Ref.~\cite{pra67e052302} proves
it in the case when $\EE_B = \HH_B$, but with a slight modification of
the proof it is easy to see that it holds for arbitrary preparable
ensemble.)  If $U_m$ denote the unitary operation corresponding to a
successful outcome $m$, then the corresponding measurement eigenstate
is necessarily
\begin{equation}
  \label{eq:badstate}
  \ket{\phi_m}_A 
  = \sqrt{p_m(\psi_0)} R^{-1} U_m^\dag \ket{\psi_0}_B.
\end{equation}

If the measurement yields the result $m$, Bob's state
becomes
\begin{equation}
  \label{eq:psii}
  \ket{\psi_m}_B = \frac1{\sqrt{p_m(\psi_0)}} RA_m \ket{\psi_0}_B.
\end{equation}
The probability $p_m (\psi_0) = \| R A_m \ket {\psi_0}_B \|^2 $ of
this outcome may depend on the target state $\ket{\psi_0}_B$ if the
protocol is not oblivious.  After Bob applied the unitary operation
$U_m$, his state must agree with the target state, so after
rearranging (\ref{eq:psii}), we have
\begin{equation}
  \label{eq:rspcond}
  U_m R A_m \ket{\psi_0}_B
  = {\sqrt{p_m(\psi_0)}} \ket{\psi_0}_B, 
  \quad \mbox{for all $\ket{\psi_0}_B \in \EE_B$}
\end{equation}
and for all outcome $m$ that does not correspond to a failure.  
The left hand side of the equation is linear in $\ket{\psi_0}_B$
since $A_m$ and $R$ are both antilinear and their product is
linear.  Therefore, the right hand side must be linear as well---as
long as the superposition states used to test its linearity are also
elements of $\EE_B$.  We show now that this implies obliviousness.

To this end, we pose an important assumption regarding the ensemble
$\EE_B$.  We will refer to it as the assumption of ``sufficiently
large ensemble.''  We assume that \emph{ $\EE_B$ consists states that
  are not linearly independent, e.g., there is a particular nontrivial
  superposition state of some other vectors that form a basis in
  $\EE_B$.}  From the practical point of view, the ensemble $\EE_B$ is
not much restricted by this assumption.  It expresses that we are not
limited to a classical alphabet but $\EE_B$ contains many states that
are evidently not all orthogonal to each other.  Essentially,
nonorthogonal states are the starting point of many quantum
information processing protocols.  For the qubit case, our assumption
is satisfied if $\EE_B$ contains at least three states.

Let $\ket k_B$ denote a (not necessarily orthogonal) basis in $\EE_B$.
Then $\ket{\psi_0}_B$ can be expanded as $\ket{\psi_0}_B = \sum_k c_k
\ket k_B$.  The linearity of the right hand side of
\eref{eq:rspcond} means:
\begin{eqnarray}
  &\sqrt{p_m(\psi_0)} \ket{\psi_0}_B =
  \sum_k c_k \left(\sqrt{p_m(k)} \ket k_B \right),\\
  &0=\sum_k c_k \left(\sqrt{p_m(k)} -\sqrt{p_m(\psi_0)} 
  \right)\ket k_B .
  \label{eq:obliv1}
\end{eqnarray}
Since the vectors $\ket k_B$ are linearly independent, only the
trivial linear combination of theirs can be zero, so
\begin{equation}
  \label{eq:obl}
  p_m(\psi_0) = p_m(k)  \quad\mbox{if $c_k \ne 0$.}
\end{equation}
Following our assumption, if there is a particular superposition state
$\ket{\psi_0^*}_B$ in $\EE_B$ such that all the $c_k$ coefficients are
nonzero, then (\ref{eq:obl}) implies that $p_m(k)$ is the same number
for every $k$.  Furthermore, $p_m(\psi_0)$ for a general
$\ket{\psi_0}_B \in \EE_B$ has no state dependence either, that is,
the protocol is oblivious.  Thus we have concluded that for an exact
RSP protocol that is capable of preparing a sufficiently large
ensemble of target states, obliviousness is equivalent to that
\emph{the measurement eigenstates $\ket{\phi_m}_A$ can be obtained
  from the target state using a fixed set of antilinear operators
  $A_m$.}

\section{Exact deterministic RSP schemes}
\label{sec:exact}

In the following, we investigate a special kind of oblivious exact
deterministic RSP protocols, and we derive a simple condition for
their existence in terms of commutation relations.

If an exact RSP protocol is deterministic (i.e., the ``failure event''
is excluded), then the condition of obliviousness given in
Sec.~\ref{sec:oblivRSP} can be reduced to a similar condition in terms
of linear operators: the scheme is oblivious if and only if the
measurement eigenstates $\ket{\phi_m}_A$ are linear functions of
the particular eigenstate $\ket{\phi_0}_A$, that is, $\ket{\phi_m}_A =
L_m \ket{\phi_0}_A$ with fixed linear operators
\begin{equation}
  \label{eq:Li}
  L_m = \sqrt{p_m(\psi_0) / p_0(\psi_0)} R^{-1} U_m^\dag U_0 R.
\end{equation}
The proof is similar to that  in Sec.~\ref{sec:oblivRSP}: the ratio
$p_m(\psi_0) / p_0(\psi_0)$ is independent of $\ket{\psi_0}_B$, then
the probabilities sum up to the probability of success that is 1.

Let us suppose first that the operators $L_m$ in (\ref{eq:Li}) are
proportional to unitaries, i.e., $L_m^\dag L_m = \ell_m \id_A$.
Then the protocol is oblivious and, exploiting that the ratio
$p_m/p_0$ is constant, we can rewrite (\ref{eq:Li}) as
\begin{equation}
  \label{eq:Li2}
  R L_m = \sqrt{p_m/p_0} U_m^\dag U_0 R.
\end{equation}
Multiplying (\ref{eq:Li2}) with its adjoint, taking its trace, and
using that $R^\dag R=\rho_A$ and $R R^\dag=\rho_B$ are the reduced
density operators of the entangled resource, we find that $\ell_m =
p_m/p_0$ and (\ref{eq:Li2}) then implies the commutation relations
\begin{equation}
  \label{eq:commute}
  [\rho_B, U_0^\dag U_m] = 0,
  \quad\mbox{for all $m$,}
\end{equation}
and similarly $ [\rho_A, L_m] = 0$.

We mention that the reverse is also true:  \emph{The operators $L_m$
  that map a particular measurement eigenstate $\ket{\phi_0}_A$ to the
  other eigenstates $\ket{\phi_m}_A$ are proportional to unitaries if
  and only if the commutation relation (\ref{eq:commute}) holds.}
Then Alice's $L_m$ and Bob's $U_m^\dag U_0$ are given in the Schmidt
bases by matrices proportional to each other.

An important implication of this theorem is that if (\ref{eq:commute})
holds for a general exact deterministic RSP scheme using partially
entangled resources, then the same set of unitaries (at Bob's side)
and the same measurement device (at Alice's side) with the same
probability distribution can be used to remotely prepare a different
ensemble using a maximally entangled resource.  To show this, we start
from the necessary and sufficient condition of RSP given in
\eref{eq:usualRSPcond}.  Multiplying it with $(D\rho_B)^{-1/2}$ from
both left and right we obtain
\begin{equation}
  \label{eq:reducedRSPcond}
  \sum_{m=0}^{N-1} p_m U_m^\dag \ket{\psi'}_{BB}\bra{\psi'} U_m
  = \frac1D \id_B, \quad \mbox{for all $\ket{\psi'}_B\in \EE_B'$,}
\end{equation}
where now the preparable ensemble $\EE_B'$ contains all the dependence on
the original partially entangled resource,
\begin{equation}
  \label{eq:EE}
  \EE_B' = \frac1{\sqrt D} U_0 \rho_B^{-1/2} U_0^\dag \EE_B
  \quad\mbox{and}\quad
  \EE_B = U_0 \sqrt{D\rho_B} U_0^\dag \EE_B'.
\end{equation}
We see that \eref{eq:reducedRSPcond} is the necessary and
sufficient condition of RSP using a maximally entangled resource.  
Note that for a maximally entangled resource the commutation relation
(\ref{eq:commute}) is trivially satisfied.

Conversely, protocols designed for maximally entangled resources can
be used with partially entangled resources if and only if
(\ref{eq:commute}) holds with the partial density operator of that
partially entangled resource.  This observation can help us construct
RSP protocols, since it is much easier to first seek for maximally
entangled schemes that satisfy (\ref{eq:reducedRSPcond}), and then
test condition (\ref{eq:commute}), while solving
(\ref{eq:usualRSPcond}) directly can be difficult.

An interesting implication of the commutation relation
(\ref{eq:commute}) is that the unitary operators $U_0^\dag U_m$ are
simultaneously block diagonal in the eigenbasis of the partial density
operator.  $\rho_B$ can always be written as a linear combination of
its eigenprojections,
\begin{equation}
  \rho_B = \sum_{j=1}^r \lambda_j P_j = 
  \bigoplus_{j=1}^r \lambda_j \id_j,
\end{equation}
where $r$ is the number of different eigenvalues of $\rho_B$, $P_j$ is
the projection onto the eigensubspace $\HH_j$, $\id_j$ is the identity
operator on $\HH_j$, and $\bigoplus$ denotes the direct sum of the
operators.  Suppose that in the lower dimensional subspaces the RSP
problem (\ref{eq:reducedRSPcond}) is solved for maximally entangled
resources.  Let us denote by $U_m^{(j)}$ the unitaries, by $p_m^{(j)}$
the probabilities, and by $\EE_j$ the preparable ensemble that
correspond to the RSP protocol in $\HH_j$,
\begin{equation}
  \label{eq:subRSPcond}
  \sum_{m=0}^{N_j-1} p_m^{(j)} {U_m^{(j)}}^\dag 
  \ket{\psi^{(j)}}_{jj}\bra{\psi^{(j)}} U_m^{(j)}
  = \frac1{D_j} \id_j, \quad \mbox{$\ket{\psi^{(j)}}_j\in \EE_j$.}
\end{equation}

Now we give a method for constructing an RSP protocol in $\HH_B$.  We
define the block diagonal unitaries as
\begin{equation}
  \label{eq:Uansatz}
  U_{k, \mathbf m} \equiv \bigoplus_{j=1}^r
  e^{2\pi ijk/r} U_{m_j}^{(j)},
\end{equation}
where $(k,\mathbf m)$ indexes an outcome of the protocol to be
constructed, with $k=1$, \ldots,~$r$ and $\mathbf m= (m_1, \ldots,
m_r)$ where $m_j$ indexes a possible measurement outcome of the RSP
protocol in the subspace $\HH_j$.  We will not necessarily consider
every combination of the $m_j$-s, but we require in our construction
that the probabilities $p_{k,\mathbf m} = p_{\mathbf m}$ do not depend
on $k$ and that 
\begin{equation}
  \label{eq:pansatz}
  r \sum_{\mathbf m} p_{\mathbf m} \delta_{nm_j} 
  = p_n^{(j)}.
\end{equation}

Then the following target states can be prepared remotely,
\begin{equation}
  \label{eq:psiansatz}
  \ket{\psi_0}_B \equiv  \bigoplus_{j=1}^r
  \sqrt{\lambda_j D_j} e^{i\varphi_j} \ket{\psi^{(j)}}_j,
  \qquad \ket{\psi^{(j)}}_j \in \EE_j,
\end{equation}
where $\varphi_j$-s are the RSP parameters that are specified by Alice
and unknown to Bob.  Neglecting the global phase factor of
$\ket{\psi_0}_B$, there are $r-1$ free of them.  If $\EE_j$ is a $d_j$
dimensional real manifold, the dimensionality of $\EE_B$ is $d=\sum_j
d_j + r-1$.

To prove that our construction indeed realizes RSP, let us substitute
$\ket{\psi'}_B = \bigoplus_{j=1}^r \sqrt{D_j/D} e^{i\varphi_j}
\ket{\psi^{(j)}}_j$ and (\ref{eq:Uansatz}) in the reduced RSP
condition (\ref{eq:reducedRSPcond}).  Verifying it block by block, we
immediately find that the off-diagonal blocks vanish because
\begin{equation}
  \label{eq:irredorto0}
  \frac1r \sum_{k=1}^re^{2\pi i(l-j)k/r} = \delta_{jl}.
\end{equation}
The RSP condition for the remaining diagonal blocks then reduces to
\eref{eq:subRSPcond} which was our starting point.

Before we present a concrete example, we note that the index $k$ is an
element of the cyclic group $\mathbb Z_r$ that is the group of integer
addition modulo $r$.  This can be straightforwardly generalized to an
arbitrary Abelian group $G$ of order $r$.  It is known from
group theory that an Abelian group $G$ is isomorphic to a direct
product of cyclic groups,
\begin{equation}
  G \simeq \mathbb Z_{r_1} \times \mathbb Z_{r_2} 
  \times \ldots \times \mathbb Z_{r_p}.
\end{equation}
Evidently, $|G| = \prod_{i=1}^p r_i = r$.  Elements of $G$ can be
treated as $p$-tuples $\mathbf k = (k_1, k_2, \ldots, k_m)$ with
$k_i=0$, 1, \ldots,~$r_i-1$.  Every element forms a conjugacy class
onto itself, therefore, all the irreducible representations of $G$ are
one dimensional, they are indexed by the group elements ($\mathbf j
\in G$), and they are products of irreducible representations of the
respective cyclic groups,
\begin{equation}
  \label{eq:irreducibles}
  u_{\mathbf j} \colon G\to \mathbb C, \quad
  u_{\mathbf j} (\mathbf k) = \prod_{i=1}^p 
  \exp(2\pi i j_i k_i /r_i).
\end{equation}
The one dimensional representations are the characters of themselves
and \eref{eq:irredorto0} can be replaced by the following
orthogonality relation of the characters of the irreducibles
\begin{equation}
  \label{eq:irredorto}
  \frac1{|G|} \sum_{\mathbf k \in G} u_{\mathbf j} (\mathbf k)
  u_{\mathbf l}^* (\mathbf k) = \delta_{\mathbf j \mathbf l},
  \qquad \mbox{for $\mathbf j, \mathbf l \in G$.}
\end{equation}

We can redefine the block diagonal unitaries (\ref{eq:Uansatz}) as
\begin{equation}
  \label{eq:UansatzAbelian}
  U_{\mathbf k, \mathbf m} \equiv \bigoplus_{\mathbf j \in G}
  u_{\mathbf j} (\mathbf k) U_{m_{\mathbf j}}^{(\mathbf j)},
\end{equation}
and we find that the preparable ensemble $\EE_B$ consists of states of
the form
\begin{equation}
  \label{eq:psiansatzAbelian}
  \ket{\psi_0}_B \equiv  \bigoplus_{\mathbf j \in G}
  \sqrt{\lambda_{\mathbf j} D_{\mathbf j}} e^{i\varphi_{\mathbf j}} 
  \ket{\psi^{(\mathbf j)}}_{\mathbf j}, \qquad
  \ket{\psi^{(\mathbf j)}}_{\mathbf j} \in \EE_{\mathbf j},
\end{equation}
where the variables $\varphi_{\mathbf j}$ are the RSP parameters.

Now we underpin our construction by a simple example.  Consider the
RSP of a qutrit state where the entangled resource has two different
eigenvalues so that $\rho_B=\diag(\lambda_1, \lambda_1, \lambda_2)$ in
the Schmidt basis.  We take the equatorial RSP protocol of
Ref.~\cite{pra63e014302} in $\HH_1$, so we have $p^{(1)}_0 = p^{(1)}_1
= \frac12$, $U^{(1)}_0 = \id$, $U^{(1)}_1 = \sigma_z$, and $\EE_1 = \{
(\ket0 + e^{i\varphi} \ket1)\sqrt2 \mid \varphi \in [0,2\pi) \}$.  For
the one dimensional subspace $\HH_2$ we take the trivial protocol
$p^{(2)}_0 = 1$, $U^{(2)}_0 = \id$, and $\EE_2 = \{ \ket2 \}$.
We construct the qutrit RSP protocol to have 4 possible outcomes
indexed by $(k,\mathbf m)$ with $k=1,2$ and $\mathbf m=(0,0), (1,0)$.
It can be easy checked that the choice $p_{k, \mathbf m} \equiv
\frac14$ for the new probabilities satisfies \eref{eq:pansatz}.  
Then \eref{eq:Uansatz} provides us with the unitary matrices
\begin{align}
  \nonumber
  U_{1,(0,0)} &= \diag (-1,-1,1),&
  U_{2,(0,0)} &= \diag (1,1,1),\\
  U_{1,(1,0)} &= \diag (-1,1,1),&
  U_{2,(1,0)} &= \diag (1,-1,1).
\end{align}
For the target states, \eref{eq:psiansatz} gives
\begin{equation}
  \ket{\psi_0}_B = \sqrt{\lambda_1} \ket0
  + \sqrt{\lambda_1} e^{i\varphi} \ket1
  + \sqrt{\lambda_2} e^{i\varphi'} \ket2,
\end{equation}
where the RSP parameter $\varphi$ originates in the RSP protocol in
$\HH_1$, while $\varphi'$ comes from our construction.  Note that it
generally holds that if we build up our protocol from the generalized
equatorial protocols of \cite{pra69e022310}, then we obtain the same
preparable ensemble as one would obtain using the protocol of
\cite{pra69e022310} directly---though with different measurement setup
and unitaries.

\section{Summary}
\label{sec:summary}

We have considered general oblivious RSP protocols.  We have presented
a necessary condition for a protocol to be oblivious: Alice's
measurement eigenstates must be antilinear functions of the target
state.  We have pointed out the importance of antilinearity by some
examples.  We have shown that our antilinear condition of
obliviousness is also sufficient if the ensemble of preparable target
states is sufficiently large.  For qubits, it means that the ensemble
contains more than two states.

We have also considered exact deterministic RSP protocols in which the
operators that map the measurement eigenstates to each other are
similar to Bob's unitaries.  We have derived conditions for the
existence of such protocols in terms of commutation relations.  We
have shown that they can be traced back to protocols using maximally
entangled resources and, therefore, our conditions are easy to use
even if we have nonmaximally entangled resources.  To underpin it, we
have constructed a protocol from subprotocols given in the
eigensubspaces of the reduced density matrix of the partially
entangled resource.

\begin{acknowledgments}
  This work was supported by the National Research Fund of Hungary
  under contracts T043287 and T049234, and by the Marie Curie
  Programme of the European Commission (MERG-CT-2004-500783).  One of
  the authors thanks A. K. Pati for useful discussions on RSP.
\end{acknowledgments}

\end{document}